\documentstyle[12pt,epsfig]{elsart}

\def\Journal#1#2#3#4{{#1} {\bf #2}, #3 (#4)}

\def\PLB{{\em Phys. Lett.}  {\bf B}}
\def\PRL{\em Phys. Rev. Lett.}
\def\PRD{{\em Phys. Rev.} {\bf D}}

\def\PRP{{\em Phys. Rep. }}
\def\EPC{{\em Eur. Phys. J.} {\bf C}}

\def\PNPP{\em Prog. Nucl. Part. Phys.}
\def\PTP{\em Prog. Theo. Phys.}

\def\JETP{\em JETP Lett.}

\def\ra{\rightarrow}
\def\be{\begin{equation}}
\def\ee{\end{equation}}

\newcommand{\mtau}{\mbox{$m_{\nu_\tau}$} }

\newcommand{\nmu}{\mbox{$\nu_\mu$}}
\newcommand{\ntau}{\mbox{$\nu_\tau$}}
\newcommand{\ds}{\mbox{$D^{\pm}_S$}}

\begin{document}
%\flushbottom
% run page numbers by "chapter"
%\input frontpages.tex
\clearpage
\title{Possible improvements on the mass of \ntau{} using leptonic $D^\pm_s$ decays}
\author{S. Pakvasa and K. Zuber}
\address{Department of Physics and Astronomy, University of Hawaii, Honolulu, HI\\
Denys Wilkinson Laboratory, University of Oxford, Keble Road, Oxford, OX1 3RH}
\maketitle
\begin{abstract}
We show how a very accurate measurement of the branching ratios of the 
leptonic
decay modes of the $D^\pm_s$ mesons can lead to an improvement in the 
mass limit for the tau neutrino.
\end{abstract}
\section{Introduction}
The last few years have seen   growing evidence for  non-vanishing neutrino 
rest masses  in the results from  
neutrino oscillation experiments \cite{zub98}. However, the direct bounds on 
neutrino masses remain rather weak.  While the electron neutrino mass is
known to be smaller than about 2.2 eV \cite{wei02}, the muon
neutrino mass has to be smaller than 170 keV
\cite{ass96}, a bound about 5 
orders of magnitude worse and still as high as 30 \% of the electron mass. The situation is even 
worse by another two orders of magnitude for \ntau{} which is known from ALEPH to be smaller than 18.2 MeV
\cite{bar98}.
In this paper we explore leptonic \ds{} decays for improving on the bound of 
\ntau.

\section{Leptonic \ds{} decays}
In the standard model the leptonic branching ratios of \ds{} are given as
\be
BR (\ds \ra l \bar{\nu}_l) = \frac{G_F^2}{8 \pi} \mid V_{cs} \mid^2 f_{D_S}^2 \tau_{D_S} m_{D_S} m^2_l F(1 - 
\frac{m_l^2}{m^2_{D_S}})^2 
\ee
with $V_{cs}$ is the corresponding CKM matrix element, $\tau_{D_S}$ the \ds{} life-time, $m_{D_S}$ the mass of the \ds,
$f_{D_S}$
the decay constant and $m_l$ the lepton mass, and F is a phase space factor 
which depends on the neutrino mass (F = 1 when neutrino mass is zero).
Several quantities cancel when  the ratio of two leptonic branching 
ratios is taken and furthermore this ratio  is quite  sensitive to the \ntau{} mass  
\cite{rek78}. Neglecting the muon neutrino mass to first approximation,
the ratio between muonic and tauonic decays can be parametrised as a function 
of $\nu_\tau$ mass to first order in $m_{\nu_\tau}$
\be
R = \frac{(\ds \ra \tau \ntau)}{(\ds \ra \mu \nmu)} = R_0 (m_{\nu} = 0) \times
\left( 1 - C (\frac{m_{\nu_\tau}}{m_\tau})^2
\right)
\ee
$R_0$ is obtained as
\be
R_0(m_\nu=0)=\frac{m^2_\tau}{m^2_\mu}\,
\frac{\left(1-\frac{m^2_\tau}{m^2_{D^\pm_s}}\right)^2}
{\left(1-\frac{m^2_\mu}{m^2_{D^\pm_s}}\right)^2}\ = 9.79
\ee
$C$ is given by
\be
\label{eq:ratio}
C = \frac{3(m_\tau /m_{\ds})^4-1}{(1-(m_\tau /m_{\ds})^2)^2} = 28.77 
\ee
The values were obtained using $m_\tau$=1777 MeV and $m_{\ds}$=1969 MeV
\cite{pdg}.
The ratio R as a function of \mtau is
plotted in Fig.1.
\begin{figure}
\begin{center}
\epsfig{file=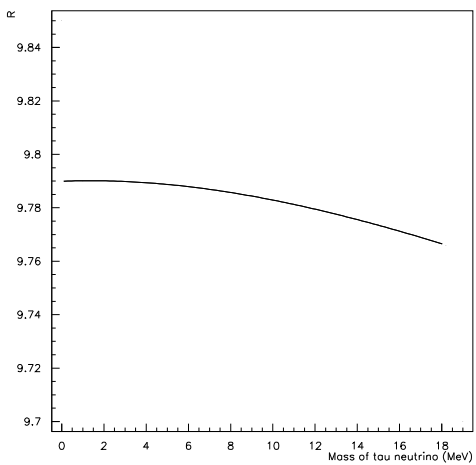,width=7cm,height=7cm}
\end{center}
\caption{The ratio R defined in Eq.(2) as a function of the mass of \ntau. 
The existing upper bound on \mtau obtained by ALEPH is 18.2 MeV.
No radiative corrections are included. For details see text.}
\end{figure} 
As expected, the figure shows that the ratio decreases as the mass of 
$\ntau$  is increased.
Furthermore, it follows from  Eq. \ref{eq:ratio} that to 
improve the bound on \mtau down to 10 MeV would correspond
to a 0.1 \% effect on $R$ only.
% and for an improvement
%down to 1 MeV would imply a 0.5 \% 
%effect only. 
To achieve the latter bound would be  especially 
interesting, because it would close the 
window for observation of a MeV Majorana neutrino in 
double beta decay via atomic mass dependent
effects\cite{hal83,zub97}.

We have not yet discussed the role of radiative corrections to the ratio $R_0$.
For the case of $\pi$ decays these have been carefully calculated and are
well-known\cite{berman,mar69}.
A complete  calculation of analogous radiative corrections for \ds{} decays is  yet to be done. It is expected that these corrections would be
in the range of 2 to 4 \%.  This will modify  the bound on $\nu_\tau$ mass. 
To extract a meaningful 
limit on the mass of $\nu_\tau$ it will be necessary to have
the radiative corrections to the value of R available.

One can also compare inclusive decay rates, rather
than the exclusive modes above. The ratio
\be
R_{\mu,\tau} = \frac{\Gamma (\ds \ra \tau \ntau + \ds \ra \tau \ntau \gamma )}{\Gamma (\ds \ra \mu \nmu + \ds \ra \mu \nmu \gamma)}
\ee
is free from any dependence on energy resolution and for a point-like
\ds{} the expression is as given in \cite{mar93}. This gives
a correction of 2.4 \%  in the direction of increasing the tau branching
ratio.  In both cases, it would be desirable to have an estimate
of the remaining structure dependent corrections, although we believe
that they would not be larger than the effects already included here. 

Similar considerations can be applied to the decays of $D^+$ as well.
However, the rates  in that case are suppressed by the CKM
suppression as well as phase space, and the branching ratios are smaller than
for $D_s$ decays by an order of magnitude or more. This make $D^+$ unsuited
for extracting mass limits on the tau neutrino.

\section{Experimental status}
The current status of leptonic branching ratios of interest are compiled in Tab. 1.
%CLEO I \cite{cleo1ds} & $D_s\rightarrow\mu$ & $344\pm37\pm52\pm42$ \\ \hline
%CLEO II\cite{cleo2ds} & $D_s\rightarrow\mu$ & $280\pm19\pm28\pm34$ \\ \hline
%E653\cite{e653} & $D_s\rightarrow\mu$ & $194\pm35\pm20\pm14$ \\ \hline
As can be seen, all measurements still have  large errors.
%Taking the PDG values \cite{pdg} of the leptonic branching ratios of \ds{} shows
%large errors on both.
%The branching ratios are quoted to be 7 $\pm$ 4 \% (for $\ds \ra \tau \ntau$) and
%0.46 $\pm$ 0.19 (for $\ds \ra \mu \nmu$).
Taking the PDG values \cite{pdg} implies R = 12.5 $\pm$ 5.5,
clearly not allowing any conclusions on \mtau. However
recently there have been improvements in investigations of \ds{} decays at LEP. 
Using the branching ratio values
of ALEPH \cite{ale00} a ratio
of R = 8.5 $\pm$ 5.2 can be obtained, unfortunately still having a much too 
large an error for the purpose at hand.
The situation might be improved  by producing a
clean and statistically large sample by looking
at diffractive \ds{} production in antineutrino-nucleon scattering at a neutrino factory \cite{del01}
or accurate measurements at the planned CLEO-c charm factory \cite{cleoc}.
For the latter an accuracy on both branching ratios of 4 \% is
predicted. If we assume that the central values remain the same as obtained by ALEPH this would imply
a value of R = 8.5 $\pm$ 0.5, which is significantly away from $R_0$. The 
accuracy in R, 
of about 6 \%, would be a great improvement over the current one, but 
still not quite at the level needed to get improved bounds 
on the $\nu_\tau$ mass, which calls for a level of less than 1 \%.   
\begin{table}
\caption{Summary of the available experimental leptonic branching
ratios of \ds . Branching ratios are given in per cent.}
%\begin{tabular}{lll}
\begin{tabular}{rcc}  
\hline
Experiment & Channel & BR \\ 
\hline
WA75\cite{wa75} & $D_s\rightarrow \mu$ & $0.4^{+0.18 +0.20}_{-0.14 -0.19}$ \\
\hline
BEATRICE\cite{beatrice} & $D_s\rightarrow\mu$ & $0.83\pm23\pm0.06\pm0.18$ \\ \hline
BES\cite{bes} & $D_s\rightarrow \mu$ & $1.5^{+1.3 + 0.3}_{-0.6 -0.3}$ \\ 
\hline
ALEPH\cite{ale00} & $D_s\rightarrow\mu$ & $0.68\pm0.11\pm0.18$ \\ \hline
\hline
L3\cite{l3} & $D_s\rightarrow\tau$ & $7.4\pm2.8\pm2.4$ \\ \hline
ALEPH\cite{ale00} & $D_s\rightarrow\tau$ & $5.79\pm0.76\pm1.78$ \\ \hline
OPAL\cite{opal} & $D_s\rightarrow\tau$ & $7.0\pm2.1\pm2.0$ \\ \hline
\end{tabular}
\end{table}

\section{Summary}
We discuss the possibility of gaining information on the mass of the tau neutrino by investigating 
leptonic \ds{} decays. An improvement on the \mtau down to 10 MeV is 
possible if the ratio of muonic and tauonic \ds{} 
decays can be measured with an 
accuracy of 0.1 \%. While current data do not allow to draw any 
conclusions , this might change in 
future experiments especially by using a charm factory. We emphasize the
importance of having a calculation of the radiative corrections available in
anticipation of a future improvement in the measurement of these branching
ratios.

\section*{Acknowledgements}
We are thankful to J. Link, W. Marciano and the referee for useful
discussions and critical comments.
K. Zuber is supported by a Heisenberg Fellowship of the Deutsche
Forschungsgemeinschaft.

\end{document}